# Biometric Blockchain: A Better Solution for the Security and Trust of Food Logistics

**Bing Xu[1], Tobechukwu Agbele[2] and Richard Jiang[2]**

[1]Computer & Information Sciences, Northumbria University, Newcastle, UK, NE1 8ST

[2]School of Computing & Communication, Lancaster University, Lancaster, UK, LA1 4WY

Correspondent e-mail: r.jiang2@lancaster.ac.uk

**Abstract.** Blockchain has been emerging as a promising technology that could totally change the landscape of data security in the coming years, particularly for data access over Internet-of-Things and cloud servers. However, blockchain itself, though secured by its protocol, does not identify who owns the data and who uses the data. Other than simply encrypting data into keys, in this paper, we proposed a protocol called Biometric Blockchain (BBC) that explicitly incorporate the biometric cues of individuals to unambiguously identify the creators and users in a blockchain-based system, particularly to address the increasing needs to secure the food logistics, following the recently widely reported incident on wrongly labelled foods that caused the death of a customer on a flight. The advantage of using BBC in the food logistics is clear: it can not only identify if the data or labels are authentic, but also clearly record who is responsible for the secured data or labels. As a result, such a BBC-based solution can great ease the difficulty to control the risks accompanying the food logistics, such as faked foods or wrong gradient labels.

## 1. Introduction

Food logistics [1], as the most important industry for everyone's daily life, is always facing great challenges in its security control. This topic has attracted a great attention by industry and researchers following a recent incident reported on BBC and Sky News that a young teenager girl, after eating a sandwich sold at the airport, died of the allergy to the peanuts gradient in the sandwich. Though it has been required by the legislation that food suppliers need to clearly label the sources and gradients in their provided foods, it is hard to monitor the whole food production processes, while multiple gradient providers may be included in one food production, and it is not rare to mislabel one food gradient with others or mistaken one source with another during the food logistics [1, 2].

As it has been widely expected, blockchain technology could provide a solution to the loose control of food logistics, and transform the entire food industry by increasing efficiency, transparency and collaboration throughout the food system [2-10]. With blockchain technology, consumers could be able to trace the source of their lettuce in seconds, shippers could see if a truck is full before they schedule a delivery, and grocery stores could verify if a carton of eggs is actually cage-free. As blockchain gets closer to its marketplace debut in the food system, it is a right time to scrutinize just how the blockchain technology will actually work in food logistics.

Blockchain was initially developed as part of the cryptocurrency, namely Bitcoin [2]. However, the technology in the cryptocurrency context looks different from how it is being developed for the food space. In the paradigm of Bitcoin, blockchain is an immutable digital ledger that works through a consensus of computer systems. Computing on the Bitcoin blockchain are essentially racing to correctly solve a calculation, and when one "wins" the race, it wins a unit of cryptocurrency and a block of data is added to the chain. The huge numbers of computer systems on the Bitcoin blockchain are why there's a huge energy cost associated with Bitcoin, a feature that would be detrimental in the agriculture space, where farmers needs to grow more and use less.

Simply in food logistics, blockchain can be just a digital ledger, a digitized record of whatever data is added by its members, with no ability to verify the accuracy of the underlying data itself. Because the truth of that data isn't actually evaluated, there's no aspect of blockchain technology that can make sure that the cage-free chicken is really cage-free or that the cabbage is actually free from pestcides.

## 2. Blockchain Technology

Blockchain [2] is a publicly-available and publicly-maintained distributed ledger (somehow equivalent to database) where all verifiable transactions are permanently recorded, and designed in such a way that data in a given block cannot be retroactively altered without altering all sequential blocks.

Originally created as the underlying technology of the Bitcoin protocol to ensure the secure transfer of the cryptocurrency (coins and tokens) without going through a central financial institution, the blockchain technology has recorded a booming success and gained the attention of technology companies doing research on the strengths, weaknesses, opportunities and threats (SWOT) as well as various applications of this technology.

It is a widely recognized that blockchain technology is a cutting-edge technology and an emerging research field with the potential to change our world like what the internet did two decades ago, once there is a strong collaboration among players in the industry. A blockchain is managed by a global peer-to-peer (P2P) network of nodes that validate new blocks using a consensus algorithm. The consensus algorithm ensures that the next block in a blockchain is the one and only version of the genuine one, thus preventing powerful adversaries from successfully branching the chain. As a result, all nodes of the network contain the same replica of data, eliminating the need of a central trusted authority to manage data [3-10].

There has been some evolution of the blockchain technology from version 1.0 to its currently version 4.0 [2-10]. Blockchain 1.0 saw the implementation of cryptocurrencies, while 2.0 was focused more on registration of smart contracts. Blockchain 3.0, on the other hand, brought about decentralized application which uses both a decentralized storage and communication. The current Blockchain 4.0 proposes more alternative approach for implementing the technology in various industries.

New application domains of the blockchain technology [2-10] include telemedicine, data markets, keyless payments, electronic voting system, and banking and financial markets. More recently, blockchain has also been seen to apply to food supply chain management, with great success on solving pending challenges in this industry area with immediate impact on commercialization.

## 3. Blockchain in Food Logistics

Though blockchain is being touted as the technology that could potentially solve challenges facing food logistics, it is yet not clear why blockchain is better than something like a database or any other form of digital information storage. Companies could simply build a database rather than build a blockchain, particularly as some of the original features of the Bitcoin version, like trustless verification, are not a generic feature.

It may be not entirely clear why blockchain is the best technology for the job of transforming the food industry, and it may or may not be. It may just be that it is the one getting attention right now, particularly as technology experts look for ways to transfer their experience and make their mark in the burgeoning food supply sector.

Where blockchain starts to reach its potential is when it is used with other technologies and systems. At the same time that the blockchain is implemented for food traceability, for example, producers can also put into place systems like enhanced water testing mechanisms or increase buffer zones between leafy green growers and livestock operations.

Blockchain can be used to gather a wealth of data and employ it in the field when used with sensors and precision delivery systems for pesticide and water all connected to a network, as with the Internet of Things. Fig.1 shows a typical framework to use blockchain in the food supply management, while data-oriented blockchain is a match of the food chain as well.

While blockchain cannot verify that an egg operation is truly cage-free or what that cage-free operation really looks like, it might offer a way for food suppliers to get more verifiable information to consumers. Food suppliers, particularly the ones who do not sell their food to the end users and cannot get an opportunity to interact with consumers, often struggle with how to engage with the customers, looking for ways to explain about details and merits the way they grow or process foods. Blockchain enables food suppliers at all stages in the supply chain to get data to consumers. That's the thing consumers can really be benefited with more information about their purchased foods.

Blockchain could be used to tell consumers that the corn was grown with pesticides, for example. It could even provide a mechanism for explaining why that pesticide is used, or a comparison of that pesticide to other pest prevention methods. The complicated nature of agriculture does not always interpret so well to a smartphone app, which might be a technologic challenge that is yet too big for blockchain to solve anyway.

To help truly understand provenance, the tracking and authentication of food supply chain is critical to finding and helping to address sources of gradients in the food supply chain worldwide. Blockchain provides a permanent record of transactions that are assembled in blocks that cannot be altered. It could serve as an alternative to traditional paper-based tracking and manual inspection systems, which can leave supply chains vulnerable to inaccuracies.

While applied to the food supply chain, digital information on food products such as farm origination details, batch numbers, factory and processing data, expiration dates, storage temperatures and shipping detail are electronically connected to each food item, and the information is entered into the blockchain along every step of the process.

The information captured in each stage in the food supply chain is agreed upon by all members of the food supply network; once there is a consensus, it becomes a permanent record that can't be altered. Each piece of information provides critical data that could potentially associate with food safety issues with the product. The records on the blockchain can also help retailers better organize the shelf-life of products in individual stores, and further strengthen safeguards related to food safety and authenticity.

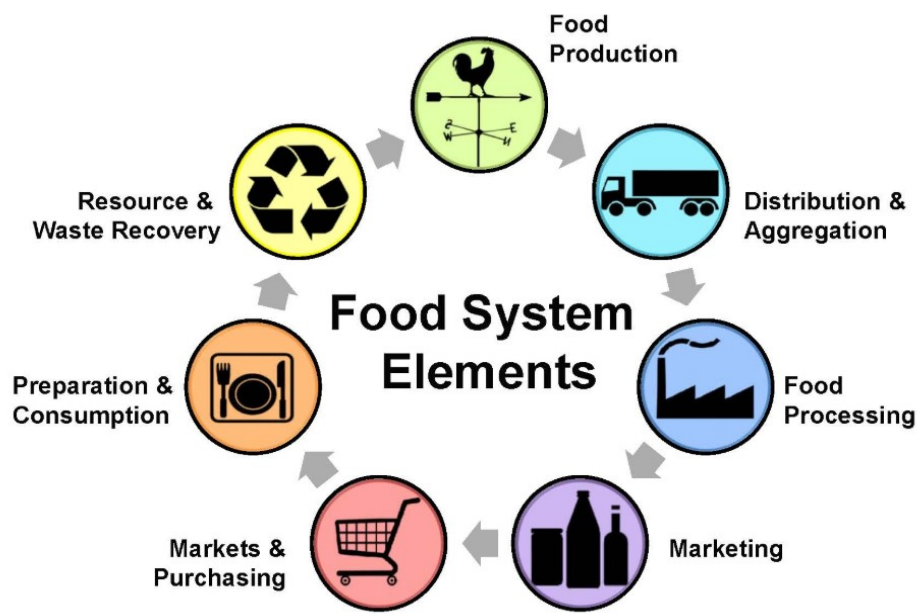

Figure 1. Blockchain can be exploited in many key stages in the food logistics [1].

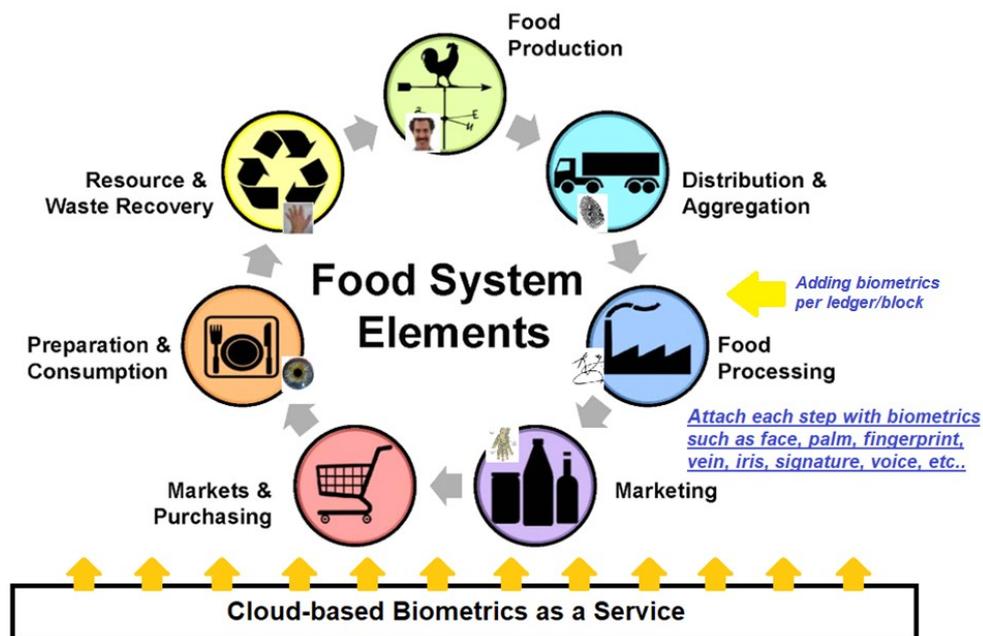

Figure 2. The proposed biometric blockchain framework for food supply chain.

Recent practical use of blockchain in food chain has been witnessed in the Walmart-IBM blockchain project. In October 2018, IBM, Walmart and Tsinghua University signed an agreement to explore food supply chain traceability and authenticity using blockchain technology in Chinese market. This food blockchain project comes as Walmart announced its new Food Safety Collaboration Center in Beijing. It is the first time to fulfil the goal in such a technology way to record how food is tracked, transported and sold to consumers across China harnessing the power of blockchain technology designed to generate transparency and efficiency in monitoring the food supply chain.

The system is yet far more restricted in scale at this initial stage. So far it is only applied to those in Walmart's leafy green food supply chain, which will likely extended into hundreds of users. As a result, the size of data to the data chain is easy to manage, with fewer verification nodes and far less energy costs overall.

With this technology, Walmart can safely tell stakeholders that a particular head of lettuce came from a particular harvest on a particular farm, so if a consumer gets sick from eating it, the police investigation will have a head start on the case. Rather than chasing a paper-based trail for months, they can immediately get to the source of a tainted head of cabbage within seconds, and that should mean a much secured world of our food supply system.

## 4. Why Biometrics is needed in Blockchain Technology

Despite the seemingly reliable and convenient services offered by the features of blockchain technology for applications, there is a host of security concerns and issues [2-10], and its understanding is relevant to the research community, governments, investors, regulators and other stakeholders.

Given the complexity and the centralised infrastructure of the blockchain, it is not a surprise that it would have some downsides especially in sharing data among different systems or regions, and securing data for a centralised infrastructure is a challenging task since potential attacks and exploits lead to a single-point-of-contact requiring trust for this individual authority. This implies that more research effort needs to be done in order to assure that information are secured in terms of privacy, assuring that only authorized users are able to access the data.

To address these challenges, recent research [11] has started to think about bring biometrics [12, 13]

into blockchain with a hope to achieve better security, scalability and privacy. Based on this initiative we proposed a new blockchain framework, namely biometric blockchain (BBC) for food logistics. Fig.2 shows our proposed framework, while each logistic step is associated with a biometric record that can be verifiable over the Biometric-as-a-Service on the cloud.

The benefits of such a BBC-based framework are apparent. First, we can easily pin down the key responsible person in each step, and make the management more transparent, controllable, and safer. Second, the food supply chain could be more robust to any attacks, such as tracing down any unknown inclusion of unwanted gradients in the food logistics. Hence, BBC as proposed in Fig.2 could be a valuable solution for practical applications in food logistics.

## 5. Privacy-Issues in Biometric Blockchain

Biometrics could secure data, and on the other side it is also sensitive to expose one's privacy. The deployment of BBC based food supply chain framework will inevitably request the biometric information from individuals, such as food makers or delivery drivers. Hence, BBC obvious needs a privacy-protected mechanism when biometric information is collected during food logistics.

Recent research has enlightened the privacy issue with a nice solution by using encrypted biometrics [14, 15, 16]. As shown in Fig.3, a set of biometric features such as faces can be encrypted and combined into a ledger as an encrypted signature. The modern technology [14, 15, 16] has revealed that we do not need to decrypt these biometric information and can directly verify it in its encrypted domain, making the use of biometrics much less sensitive in term of privacy concerns when it is used for blockchain technology.

The implementation of BBC may be associated with the online biometric verification, and may be based on multimodal biometrics [17]. Classically, feature (such as LBP or SIFT) based [18, 19] biometric verification is popular though it could be a bit time-consuming. Recent approaches such as deep neural networks [20, 21, 22] are seen gradually taking over the domain of biometric verification.

There may rise concerns on the computing resources while biometrics-as-a-service is accessed via cloud platforms, as shown in Fig.2. However, in a BBC-based food logistic chain, biometric verification may happen only for necessary check, which means those biometric information could mostly be dormant and hence the requirement on extra computing could be minimized.

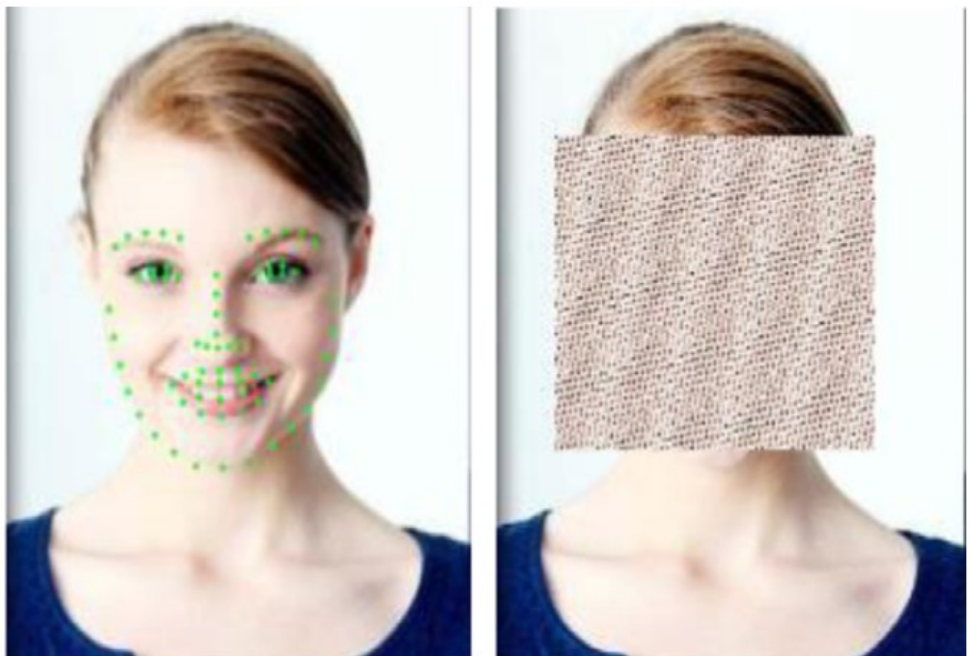

Figure 3. Privacy protection using encrypted biometrics in blockchain [14].

## 6. Conclusion

Blockchain emerging as a promising technology could totally change the landscape of data security in the coming years, especially for the data access over Internet-of-Things and cloud servers. However, blockchain itself does not contain any information about who owns the data and who uses the data. In this paper, other than simply encrypting data into keys, we proposed a new protocol called Biometric Blockchain (BBC), in which the biometric cues of individuals are explicitly incorporated in the BBC protocol to unambiguously identify creators or users in a blockchain-based system. Such a new protocol can particularly address the increasing needs to secure the food logistics, following the recently widely reported incident on wrongly labelled foods that caused the death of a passenger on airplane. The merits of using BBC in the food logistics are visibly clear: it can not only identify if the data or labels are authentic, but also clearly record who is responsible for the secured data or labels. Hence, such a BBC-based solution can greatly compromise the difficulty to control the critical issues in food logistics, such as food authenticity and gradient mislabelling.